\documentclass[aps,twocolumn,superscriptaddress,showkeys,showpacs]{revtex4-1}
\usepackage{amssymb}
\usepackage{amsfonts}
\usepackage{amsmath}
\usepackage{amsthm}
\usepackage{epsfig}
\usepackage{color} 
\usepackage{subfigure}

\begin{document}

\title{Transition from spatial coherence to incoherence in coupled chaotic systems}

\author{Iryna Omelchenko}
\affiliation{Institut f{\"u}r Theoretische Physik, TU Berlin, Hardenbergstra\ss{}e 36, 10623 Berlin, Germany}
\affiliation{Bernstein Center for Computational Neuroscience, Humboldt-Universit{\"a}t zu Berlin, Philippstra{\ss}e 13, 10115 Berlin, Germany}
\author{Bruno Riemenschneider}
\affiliation{Institut f{\"u}r Theoretische Physik, TU Berlin, Hardenbergstra\ss{}e 36, 10623 Berlin, Germany}
\author{Philipp H{\"o}vel} 
\affiliation{Institut f{\"u}r Theoretische Physik, TU Berlin, Hardenbergstra\ss{}e 36, 10623 Berlin, Germany}
\affiliation{Bernstein Center for Computational Neuroscience, Humboldt-Universit{\"a}t zu Berlin, Philippstra{\ss}e 13, 10115 Berlin, Germany}
\author{Yuri Maistrenko}
\affiliation{Institute of Mathematics, National Academy of Sciences of Ukraine, Tereshchenkivska str. 3, 01601 Kyiv,
Ukraine}
\affiliation{National Center for Medical and Biotechnical Research, National Academy of Sciences of Ukraine,
Volodymyrska str.54, 01030 Kyiv, Ukraine}
\author{Eckehard Sch{\"o}ll}
\email[corresponding author: ]{schoell@physik.tu-berlin.de}
\affiliation{Institut f{\"u}r Theoretische Physik, TU Berlin, Hardenbergstra\ss{}e 36, 10623 Berlin, Germany}

\date{\today}

\begin{abstract}
We investigate the spatio-temporal dynamics of coupled chaotic systems with nonlocal interactions, where each element is
coupled to its nearest neighbors within a finite range. Depending upon the coupling strength and coupling radius, we
find characteristic spatial patterns such as wave-like profiles and study the transition from coherence to incoherence
leading to spatial chaos. We analyze the origin of this transition based on numerical simulations and support the
results by theoretical derivations identifying a critical coupling strength and a scaling relation of the coherent
profiles. To demonstrate the universality of our findings we consider time-discrete as well as time-continuous chaotic
models realized as logistic map and R\"ossler or Lorenz system, respectively. Thereby we establish the
coherence-incoherence transition in networks of coupled identical oscillators.
\end{abstract}


\maketitle

\section{Introduction}
While isolated nonlinear systems are fairly well understood, much less is known about emergent dynamics of coupled
systems \cite{KUR84,KAN96a,MOS02,PIK03,SCH07,NAK10}. Complete synchronization, also known as zero-lag synchronization,
where all nodes
of a network behave identically, is just the simplest scenario. Nevertheless, it has important applications, for
instance, as chaos synchronization for secure communications \cite{SHA06,FIS06}. Other types of synchronization have
been found as well, including phase-lag, cluster, or generalized synchronization \cite{PIK03}. Recently a new form of
dynamics called chimera state triggered a lot of research
\cite{KUR02a,ABR04,ABR06a,OME08,LAI09,LAI09a,WOL11,WOL11a,LEE11,LAI11,SIN11}. These chimera states arise in regular networks
of identical
phase oscillators and consist of parts with high spatial coherence and regions where spatial coherence is lost. They can be realized
with carefully prepared initial conditions and a nonlocal network topology, which both are crucial ingredients for
these states. Similar phenomena, localized patterns surrounded by a chaotic background, were found in populations of threshold elements with nonlocal interaction, related to
a continuity-discontinuity transition~\cite{SAK98}. 

In this paper we investigate in detail the transition from spatial coherence to incoherence in both time-discrete and
time-continuous chaotic systems, using the well-known models of the logistic map and the R\"ossler system, respectively. See also Ref.~\cite{OME11} for a first report on this transition. We consider a ring topology and use the coupling strength and coupling range as bifurcation parameters. At the transition we observe partially incoherent chimera-like states that we identify as spatial chaos \cite{CHO95a,NIZ02}.

This paper is structured as follows: Section~\ref{sec:models} introduces the model equations. Sections~\ref{sec:diagram} and \ref{sec:transition} describe the general coherence-incoherence scenario and numerical results on the coherence-incoherence transition, respectively. Analytical results are presented in Sec.~\ref{sec:analytical}. 
In Sec.~\ref{sec:lorenz} we demonstrate that the observed scenario can also be found in the Lorenz model.
Finally we conclude in Sec.~\ref{sec:conclusion}.

\section{Models}\label{sec:models}
In order to investigate the universality of the coherence-incoherence transition for $N$ coupled oscillators we consider different nonlinear models: As chaotic systems we employ the time-discrete, logistic map as well as the R\"ossler system. The latter is paradigmatic for chaos described by ordinary differential equations.

As a control parameter we use the coupling strength $\sigma$ and the coupling radius $r=P/N$, where $P$ denotes the
number of nearest neighbors in each direction on a ring. Varying the value of $P$ from $1$ to $N/2$ we cover the range
of regular network topologies from local (nearest-neighbor) to global (all-to-all) coupling.

\subsection{Logistic map}\label{sub:map}
As an example for time-discrete systems, we study coupled maps:
\begin{equation}\label{eq:map}
z_i^{t+1} = f\left(z_i^t\right) + \dfrac{\sigma}{2P} \sum\limits_{j=i-P}^{i+P} \left[ f\left(z_j^t\right) -
f\left(z_i^t\right) \right],
\end{equation}
where $z_i$ ($i=1,...,N$) are real dynamic variables and $t$ denotes the discrete time. In the coupling term
the index $i$ is periodic mod $N$. The function $f(z)$ is a local one-dimensional map that we choose as the logistic
map $f(z)=az(1-z)$. We fix the bifurcation parameter $a$ at the value $a=3.8$ unless stated otherwise. This is a value
that yields chaotic behavior with a positive Lyapunov exponent $\lambda \approx 0.431$.

\subsection{R\"ossler system}\label{sub:roessler}
In order to show that the coherence-incoherence transition scenario is universal, we will also discuss nonlocally coupled networks of time-continuous systems. As a prominent example for chaotic systems, we consider the R{\"o}ssler model. In analogy to Eq.~(\ref{eq:map}) the compound system is given by
\begin{equation}\label{eq:roessler}
\begin{array}{l}
\dot{x}_i = -y_i-z_i + \dfrac{\sigma}{2P} \sum\limits_{j=i-P}^{i+P} \left( x_j - x_i \right), \\
\dot{y}_i = x_i + ay_i+ \dfrac{\sigma}{2P} \sum\limits_{j=i-P}^{i+P} \left( y_j - y_i \right), \\
\dot{z}_i = b+z_i(x_i-c)+\dfrac{\sigma}{2P} \sum\limits_{j=i-P}^{i+P} \left( z_j - z_i \right), 
\end{array}
\end{equation}
with $i =1,...,N$. Again the index $i$ is periodic mod $N$ to realize the
underlying ring topology. The system's parameters are chosen as $a=0.42$, $b=2$, and $c=4$ showing chaotic
behavior in the uncoupled case. 

The $3N$ coupled Eqs.~(\ref{eq:roessler}) can also be written in compact form as follows
\begin{equation}\label{eq:roessler2}
\dot{{\bf X}} = {\bf F}({\bf X}) + \frac{\sigma}{2P} ({\bf G} \otimes {\bf H}){\bf X}
\end{equation}
with ${\bf X} = (x_1,y_1,z_1,\dots,x_N,y_N,z_N)$ and a $3\times3$ coupling matrix ${\bf H}$ that determines, which
components are transmitted. The $N\times N$ matrix ${\bf G}$ describes the network topology.
In the case of Eqs.~(\ref{eq:roessler}), we have the identity matrix for ${\bf H}$, and a circulant matrix with
cyclically permutated rows $(-2P,\underbrace{1,\dots,1}_{P},0,\dots,0,\underbrace{1,\dots,1}_{P})$ is chosen for ${\bf
G}$ such that $g_{ii}=-2P$.

This notation has been employed in the framework of the master stability function in order to investigate complete, potentially chaotic synchronisation \cite{PEC98}. Based on a variational equations this formalism allows for calculation of the Lyapunov exponent transverse to the synchronisation manifold \cite{CHA05,HWA05}. The master stability function technique has been used in various contexts ranging from generic models \cite{CHO09,FLU10b} to coupled lasers \cite{HEI11,FLU11b} and neuroscience \cite{DHA04,DHA04a,LEH11}.
It can also be extended to cluster synchronization \cite{SOR07}, but does not give conclusive results for the purpose of the
investigation of spatially incoherent states.

\section{Structure of coherence tongues}\label{sec:diagram}

\begin{figure}[ht!]
\includegraphics[width=0.9\linewidth]{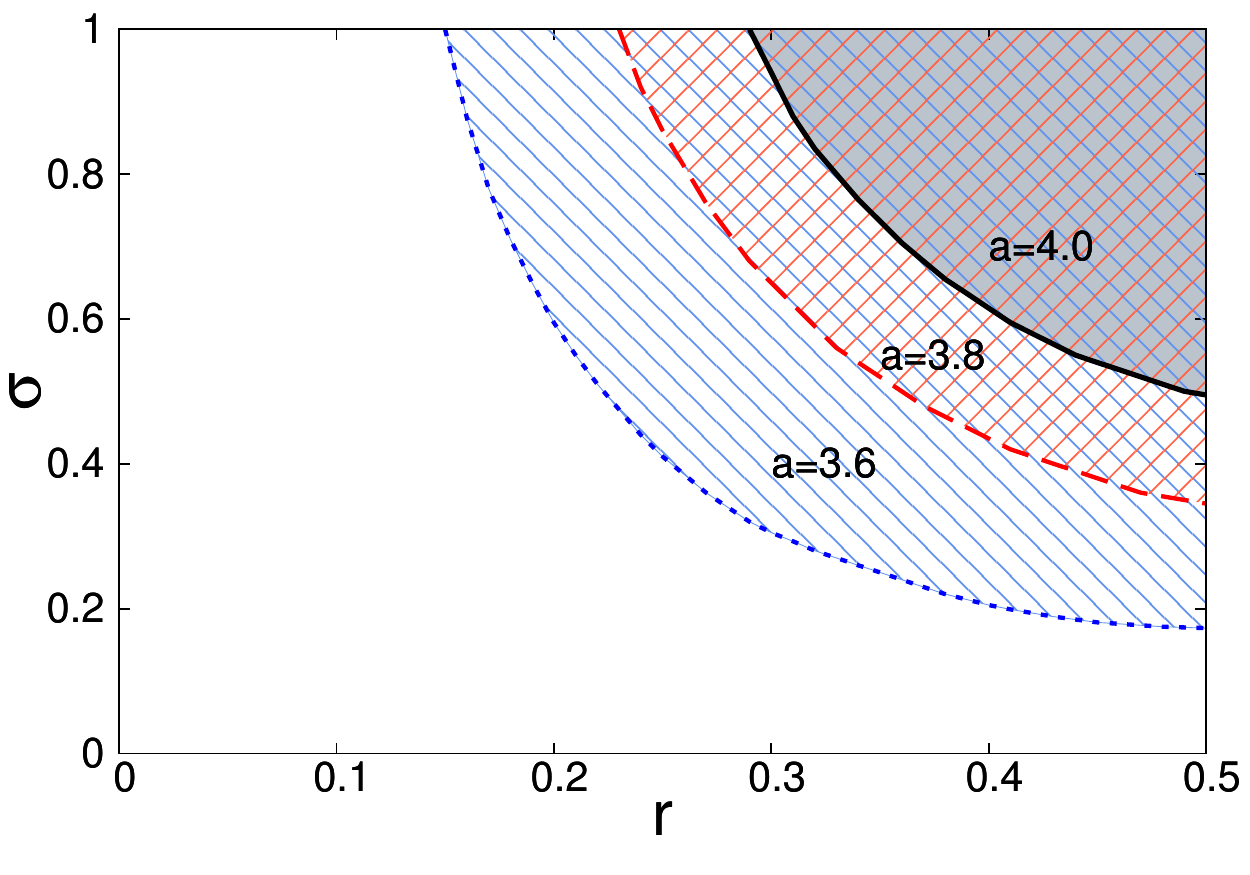}
\caption{(Color online) Blowout bifurcation curves in the $(r,\sigma)$-parameter plane for $N=100$ coupled logistic
maps. The system parameter is chosen as $a=4.0$, $3.8$, and $3.6$ for the solid, dashed (red), and dotted (blue) curve,
respectively.}
\label{fig:BB}
\end{figure}

Let us discuss first the coherence-incoherence transition for coupled logistic maps described by Eq.~(\ref{eq:map}). For large values of coupling radius $r$ and coupling strength $\sigma$ we observe complete chaotic synchronization.
There the oscillators behave identically ($z_{sync} = z_1=...=z_N$), but chaotically in time following the dynamics of $f$. 
Using the fact that the coupling matrix  of system~(\ref{eq:map}) is a circulant matrix, with cyclically permutated rows
$(-2P,\underbrace{1,...,1}_{P}, 0,...,0,\underbrace{1,...,1}_{P}),$
we obtain analytically the values for $N-1$ transversal Lyapunov exponents of the completely synchronized solution
$z_{sync}$: 
\begin{equation*}
\begin{array}{rl}
\lambda_{\bot,m} =
\lambda_{\parallel} + & \ln 
\left|
(1-\sigma) + \dfrac{\sigma}{2P}
\left(
\sum\limits_{k=1}^{P} \cos \left(\dfrac{2 \pi m k}{N} \right)\right.\right. \\
 & \left.\left. + \sum\limits_{k=N-P}^{N-1} \cos \left(\dfrac{2 \pi m k}{N} \right)\right)\right|,
\end{array}
\end{equation*}
$m=1,...,N-1$, where $\lambda_{\parallel}>0$ denotes the longitudinal Lyapunov exponent corresponding to the
one-dimensional local map, and the second term on the right-hand side is derived from the eigenvalues of the Jacobian
matrix of the system, using the fact that coupling matrix is circulant.

With a decrease of the coupling radius $r$ or the coupling strength $\sigma$ the completely synchronized solution
loses its stability in a blowout bifurcation, where the largest transverse Lyapunov exponent becomes positive. The
position of the blowout bifurcation line in the $(r,\sigma)$-plane depends on the local dynamics of the system elements.
Figure~\ref{fig:BB} shows examples for the blowout bifurcation curves for $a=3.6$, $3.8$, and $4.0$. For any of these
values the local logistic maps operate in the chaotic regime. The region of complete chaotic synchronization is the region  
above the blowout bifurcation curves.

The completely synchronized state represents the simplest example of spatial coherence. 
To be precise, we will call a network state $z_i^t$, $i=1,...,N$, coherent on the ring $\mathcal{S}^1$ as $N
\rightarrow \infty$,  if for any point $x\in\mathcal{S}^1$ the following limit holds:
\begin{equation}
\lim\limits_{N \rightarrow \infty} \lim\limits_{t \rightarrow \infty} \sup\limits_{i,j \in U_{\delta}^N (x)} \left|
z_i^t - z_j^t \right|\rightarrow 0, \quad \text{for} \quad \delta\rightarrow 0,
\label{Eq:CoherenceDef}
\end{equation}
where $U_{\delta}^N (x) = \left\{ j:~0 \leq j \leq N,~ \left| {j}/{N} -x \right| < \delta \right\}$ denotes a network
neighborhood of the point $x$. Intuitively it means that the profile becomes smooth in the thermodynamic limit $N
\rightarrow \infty$. If the limit~(\ref{Eq:CoherenceDef}) does not vanish for $\delta\rightarrow 0$, at least for one
point $x$, the network state is considered incoherent.  

\begin{figure}[ht!]
\includegraphics[width=\linewidth]{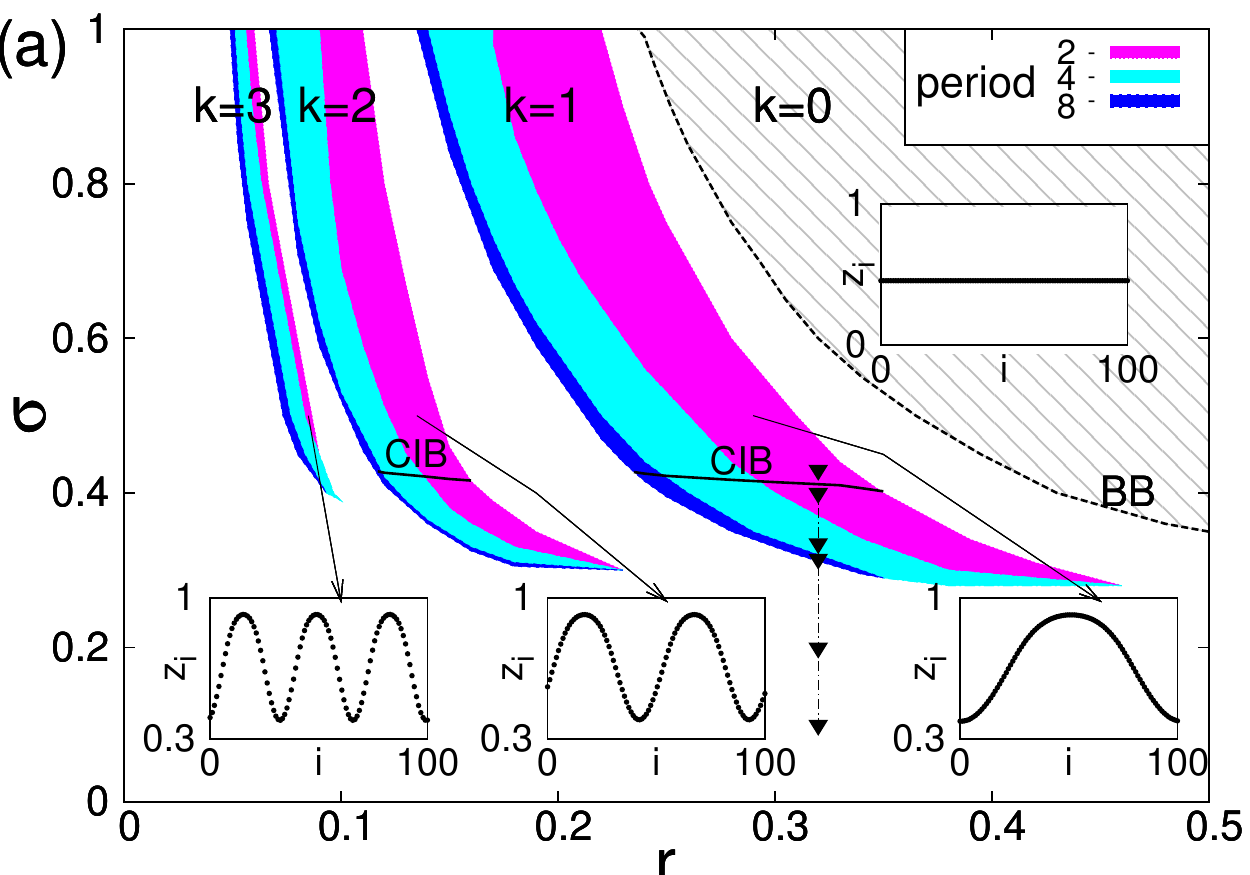}
\includegraphics[width=\linewidth]{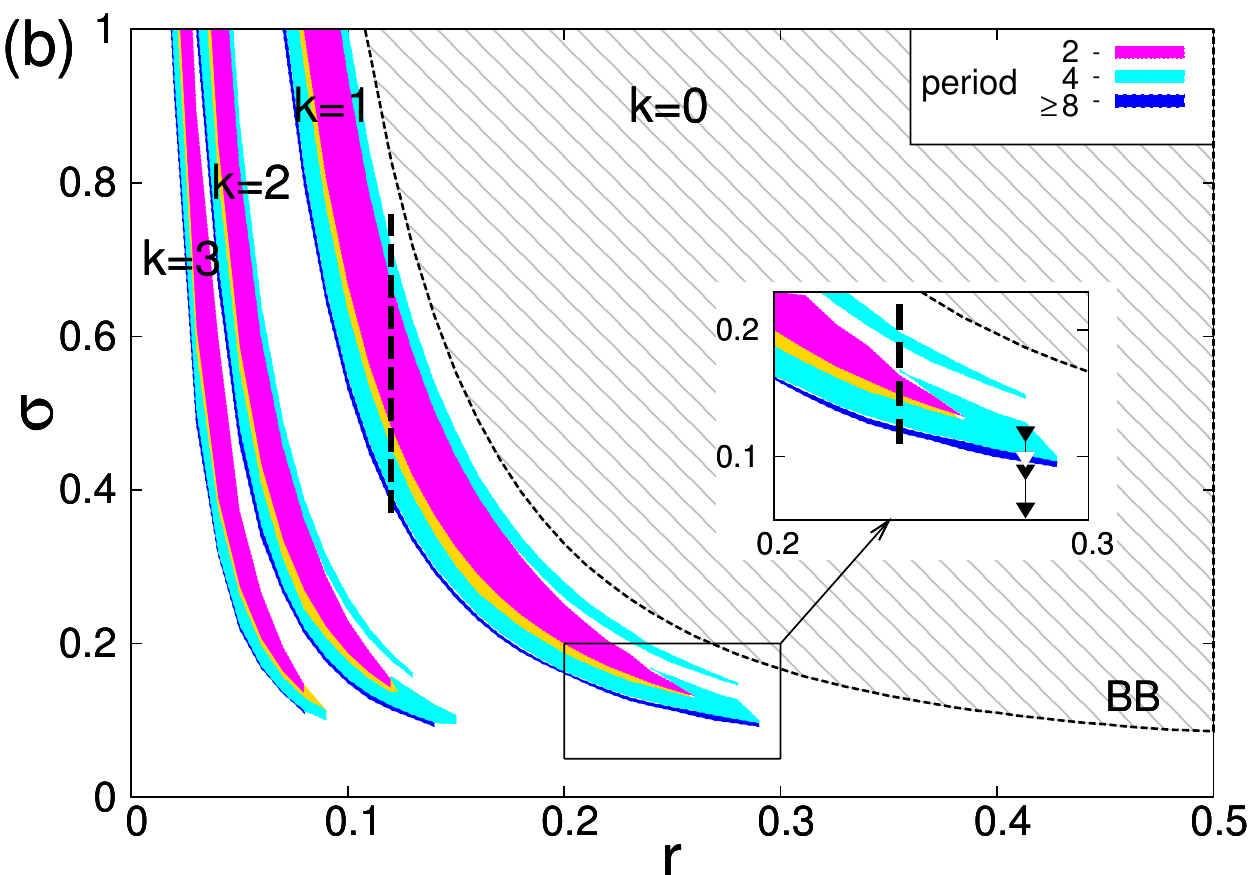}
\caption{(Color online) Coherence regions in the $(r,\sigma)$ parameter plane for $N=100$ logistic maps (a) and
R\"ossler
systems (b), labeled by the wave number $k$. Gray scale (color code) inside the coherence regions
distinguishes different periods and the coherence-incoherence bifurcation (CIB) curve separates regimes with coherent
and incoherent dynamics.
The light hatched region bounded by the blowout bifurcation curve BB refers to completely synchronized chaotic states.
The insets in panel (a) display snapshots of typical coherent states. System parameters: panel (a) $a=3.8$; panel (b)
$a=0.42$, $b=2$, and $c=4.0$. The vertical dashed lines refer to values of $\sigma$ in
Fig.~\ref{fig:sweep}. Triangles denote parameter values used to describe the bifurcation scenario in
Figures~\ref{fig:maps_snap} and~\ref{fig:RoesslerBifScenario}.}
\label{fig:diagram}
\end{figure}

Below the blowout bifurcation curve, several further regions of coherence appear as the control parameters $\sigma$
or $r$ decrease. Figure~\ref{fig:diagram} depicts these regions as shaded (color) tongues for the logistic map and
the R\"ossler
system in panels (a) and (b), respectively. The blowout bifurcation line is added as dashed curve (BB), and the regime
of complete, chaotic synchronisation is indicated by the light hatching.

A coherent state has a smooth profile characterized by the number of maxima, i.e., the wave number $k$. Only regions for
wave numbers $k = 1$, $2$, and $3$ are shown. The corresponding coherent states $z_i^t$ are shown as snapshots in the
insets of
Fig.~\ref{fig:diagram}(a) for the logistic map. Further decrease of $r$ yields additional thin higher-order regions
following a period-adding cascade $k =4,5,\dots$ (not shown). Inside the regions, the states are coherent in space and
periodic in
time and undergo a period-doubling cascade in time as $r$ or $\sigma$ decrease. In the parameter space
between the coherence regions the network dynamics remains coherent but not periodic anymore. The states alternate
chaotically between the adjacent $k$-states and thus exhibit chaotic itineracy~\cite{KAN96a,KAN03}. The combination of
period-adding in space and period-doubling in time represents a remarkable feature of networks of coupled
chaotic oscillators with nonlocal coupling.

Figure~\ref{fig:diagram}(b) shows stability diagram for the coupled R\"ossler systems again in the $(r,\sigma)$-plane.
It is easy to see that the overall structure of the diagram is similar to the one obtained for nonlocally coupled
logistic maps
(Fig.~\ref{fig:diagram}(a)). In analogy to the maps, there exist a blowout bifurcation boundary (BB curve) and a
sequence of coherence tongues with wave numbers $k$ indicating the universality of the general dynamical scenario.
In the shaded (color) regions shown in Fig.~\ref{fig:diagram}(b) the coherent states with regular (periodic or
quasiperiodic) dynamics are the only stable solution. The
bright (yellow) regions inside the coherence tongues below the stable period-2 tongues refer to multistability and more complex behavior including
coexisting period-4 and torus solutions in between the period-2 and period-4 regions.
The dark (blue) region is the region of stability for the solutions with periods $8$ and larger. Since these regions
appear to be very thin, we combine them for better visualization. 

\begin{figure}[ht!]
\includegraphics[width=\linewidth]{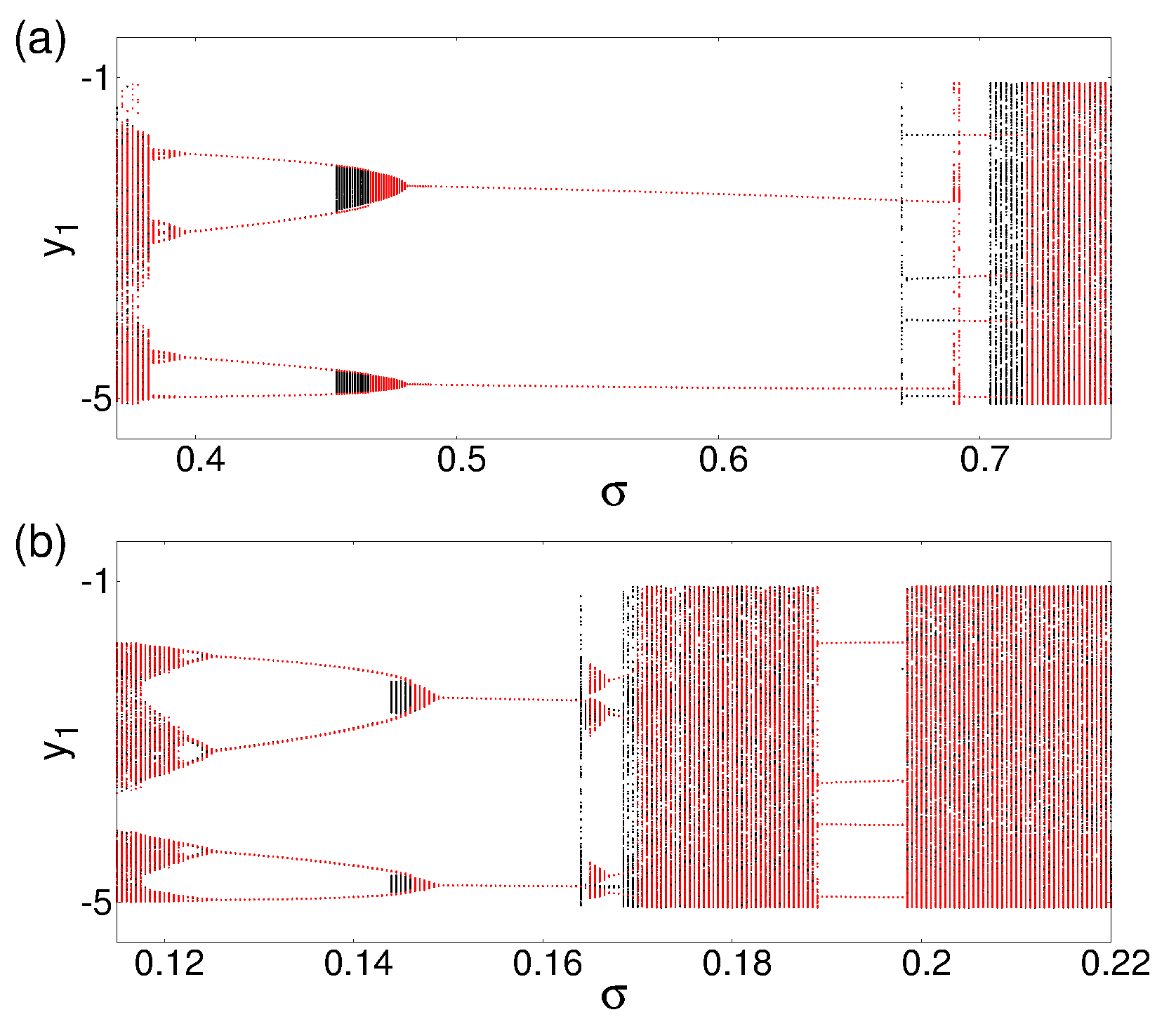}
\caption{(Color online) Bifurcation scenario at cuts through the $k=1$ coherence tongue of Fig.~\ref{fig:diagram}(b) at (a) $r=0.12$ and (b) $r=0.24$. The Poincar\'{e} section ($x_1=0$) of $y_1$ is shown for the R\"ossler systems in dependence on the coupling strength $\sigma$. The black and gray (red) points correspond to a slow down and up sweep of $\sigma$, respectively. Long transients up to 3000 oscillation periods have been discarded.
Other parameters as in Fig.~\ref{fig:diagram}.
}
\label{fig:sweep}
\end{figure}

For further insight of the transitions within the coherence tongues Figs.~\ref{fig:sweep}(a) and (b)
depict cuts through the $(r,\sigma)$-parameter space for a fixed coupling radius $r=0.12$
and $0.24$, respectively (see dashed lines in Fig.~\ref{fig:diagram}(b)). The black and gray (red) values refer to the
variable $y_1$ at a Poincar\'{e} section ($x_1=0$) for a very slow decrease and increase of the coupling strength
$\sigma$ with steps $\Delta\sigma=0.001$, respectively. For each value of $\sigma$ 300 iterations are shown, where each
initial condition is chosen as the final configuration of the previous value of $\sigma$. Panel (a) shows a sequence of
chaotic, period-4, torus, period-2, coexisting period-2 and period-4, period-4, and again chaotic dynamics. In panel (b)
there appears an additional regime of chaos in between the period-2 and period-4 region.


The coupling radius appears to be the crucial parameter in the system with nonlocal couplings. With increasing the
system size $N$ and a proportional increase of the number of coupled neighbors, we obtain similar coherent solutions. If
with increasing $N$ the number of coupled neighbors $P$ does not change, then the coupling radius decreases, and the
point of operation $(r,\sigma)$ moves in the stability diagram (Fig.~\ref{fig:diagram}) to the left, obtaining coherent
solutions with larger wavme numbers or itineracy regions between them.    

The bifurcation diagrams presented in Fig.~\ref{fig:diagram} have been obtained by means of numerical
simulations. Note, however, that in the thermodynamic limit $N\rightarrow\infty$ it is also possible to obtain some
analytic results. Details on such derivations are given in Sec.~\ref{sec:analytical}, e.g., by identifying a scaling
relation of the coherent profiles.

\section{Coherence-incoherence transition}\label{sec:transition}

\begin{figure}[ht!]
\includegraphics[width=\linewidth]{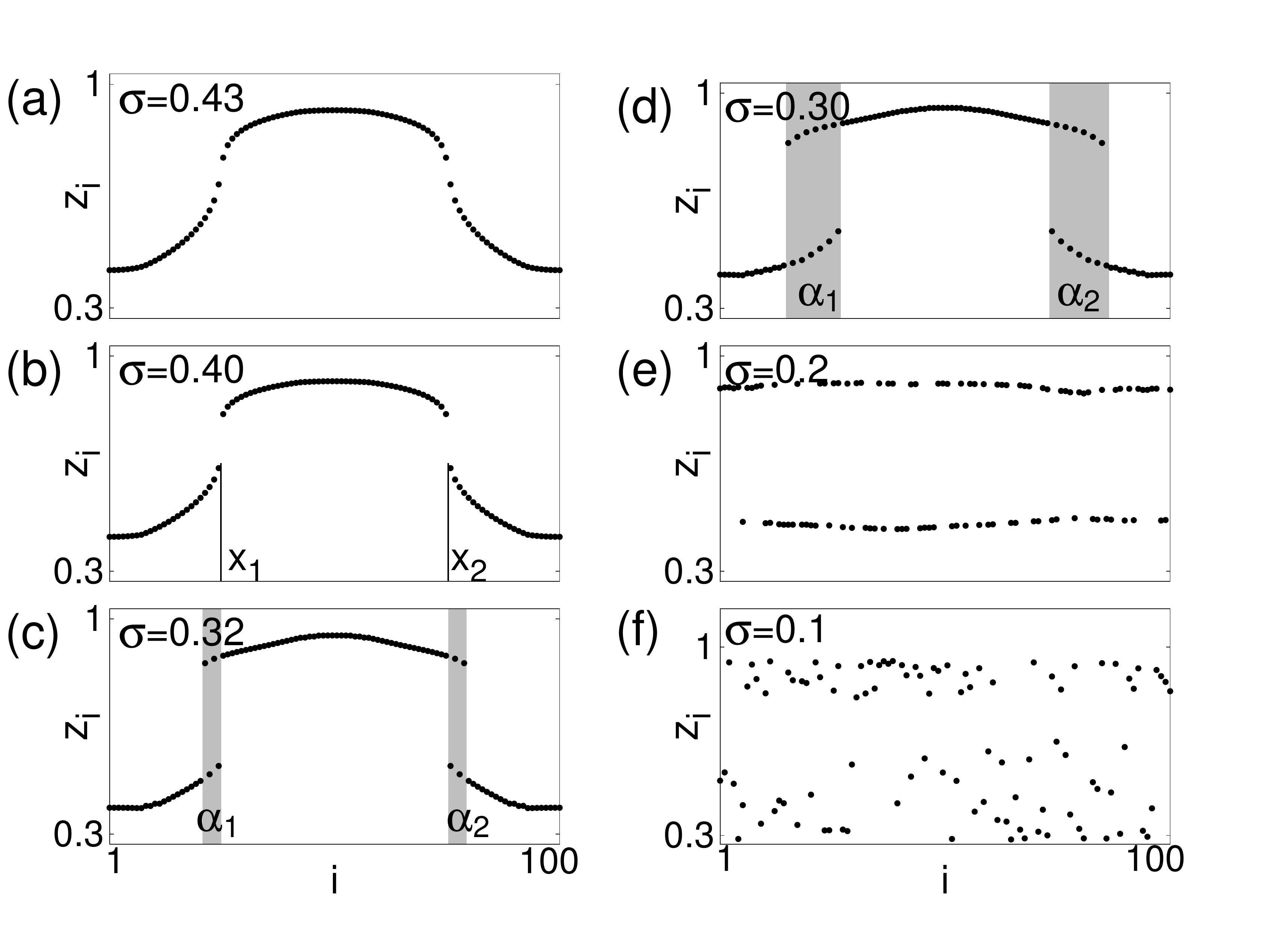}
\caption{Snapshots of the dynamics of the logistic map for different coupling strengths $\sigma= 0.43, 0.4,
0.32, 0.3, 0.2,$ and $0.1$. The gray regions labeled $\alpha_1$ and $\alpha_2$ refer to multistable
incoherent parts of the profile. The coupling radius is fixed at $r=0.32$. Other parameters as in
Fig.~\ref{fig:diagram}.}
\label{fig:maps_snap}
\end{figure}

The described combination of period-adding in space and period-doubling in time for coherent solutions is valid for
sufficiently large coupling strength ($\sigma \gtrsim 0.4$). In Ref.~\cite{OME11} we have shown that with decreasing
coupling strength $\sigma$ a coherence-incoherence bifurcation is observed in the
time-discrete system~(\ref{eq:map}). The typical scenario of this transition is shown in Fig.~\ref{fig:maps_snap} for
the logistic map: The smooth wave-like solution profile $z_i^t$ splits up into upper and lower branches. Furthermore,
two narrow boundary layers of incoherence appear around the points $x_1$ and $x_2$ in Fig.~\ref{fig:maps_snap}(b). They
are indicated by shaded stripes $\alpha_1$ and $\alpha_2$ in
Figs.~\ref{fig:maps_snap}(c) and (d). The width of the incoherence stripes increases for smaller coupling strength $\sigma$ (see Fig.~\ref{fig:maps_snap}(d)) until the dynamics becomes completely incoherent as depicted in Figs.~\ref{fig:maps_snap}(e) and (f).

Below the bifurcation parameter value $\sigma \approx 0.40$ numerous hybrid, that is, partially coherent states arise.
They are characterized by coherence on some intervals of the ring $\mathcal{S}^1$ and incoherence on the complementary
intervals. Figs.~\ref{fig:maps_snap}(c) and (d) display exemplary realizations of these partially coherent states.
Within the incoherence intervals, i.e., $\alpha_1$ and $\alpha_2$, any combinations of the upper and lower states are
admissible for appropriately prepared initial conditions. 

Thus the coherence-incoherence bifurcation leads to spatial chaos \cite{CHO95a,NIZ02} that appears first at narrow
incoherence intervals and, with decreasing coupling strength, encompasses the whole ring. Therefore we identify a
chimera-like state of coexisting coherent and incoherent regions as a transitional state in the coherence-incoherence
bifurcation scenario. Opposed to previously reported chimera states in time-continuous systems
\cite{KUR02a,ABR04,SET08,LAI10,OME10a}, however, the temporal behavior in the present case is periodic rather than
chaotic. The complexity arises due to the huge variety of multistable incoherent states and corresponds to permutations
of the sequence of upper and lower local states. With further decrease of the coupling strength, the chimera states
cease to exist and we find completely incoherent behavior.

The findings above are based on numerical simulations of coupled chaotic logistic maps. Even if the local logistic maps operate in the periodic regime, one can observe a sequence similar to Fig.~\ref{fig:maps_snap} as reported in Ref.~\cite{OME11}. Note that, when the local dynamics of the system is periodic, the region of stability of the completely synchronized solution covers almost all of the  $(r,\sigma)$ parameter plane, and the system is characterized by multistability.

\begin{figure}[ht!]
\includegraphics[width=\linewidth]{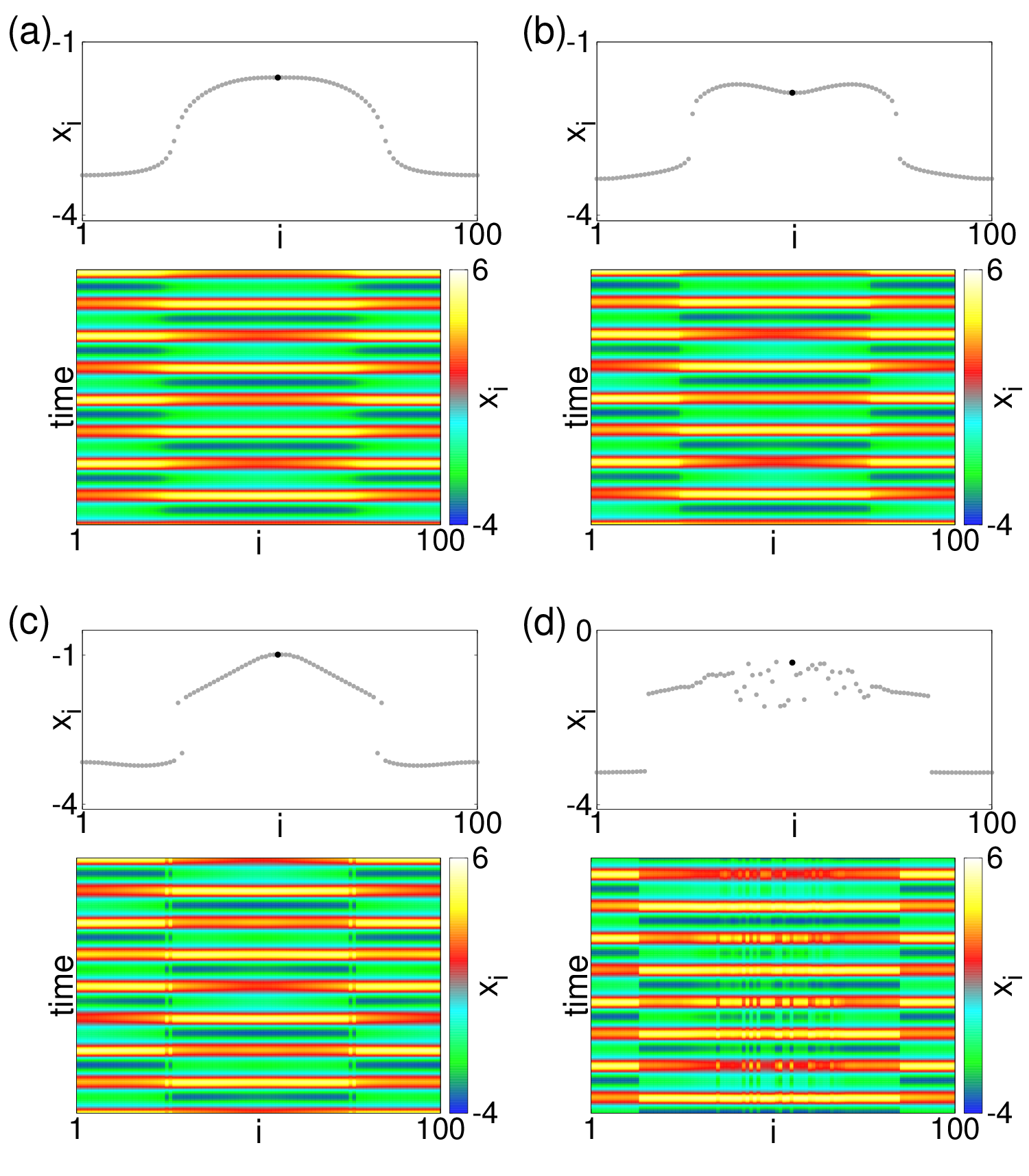}
\caption{(Color online) Snapshots of the variables $x_i$ (upper panels, $x_{50}$ as black dot) and space-time
plots (lower panels) for the R\"ossler system. The coupling strength is chosen as $\sigma= 0.12, 0.1, 0.09$, and $0.06$
in (a) to (d), respectively. The coupling radius is fixed at $r=0.28$. Other parameters as in Fig.~\ref{fig:diagram}.}
\label{fig:RoesslerBifScenario}
\end{figure}

\begin{figure*}[ht!]
\includegraphics[width=0.8\linewidth]{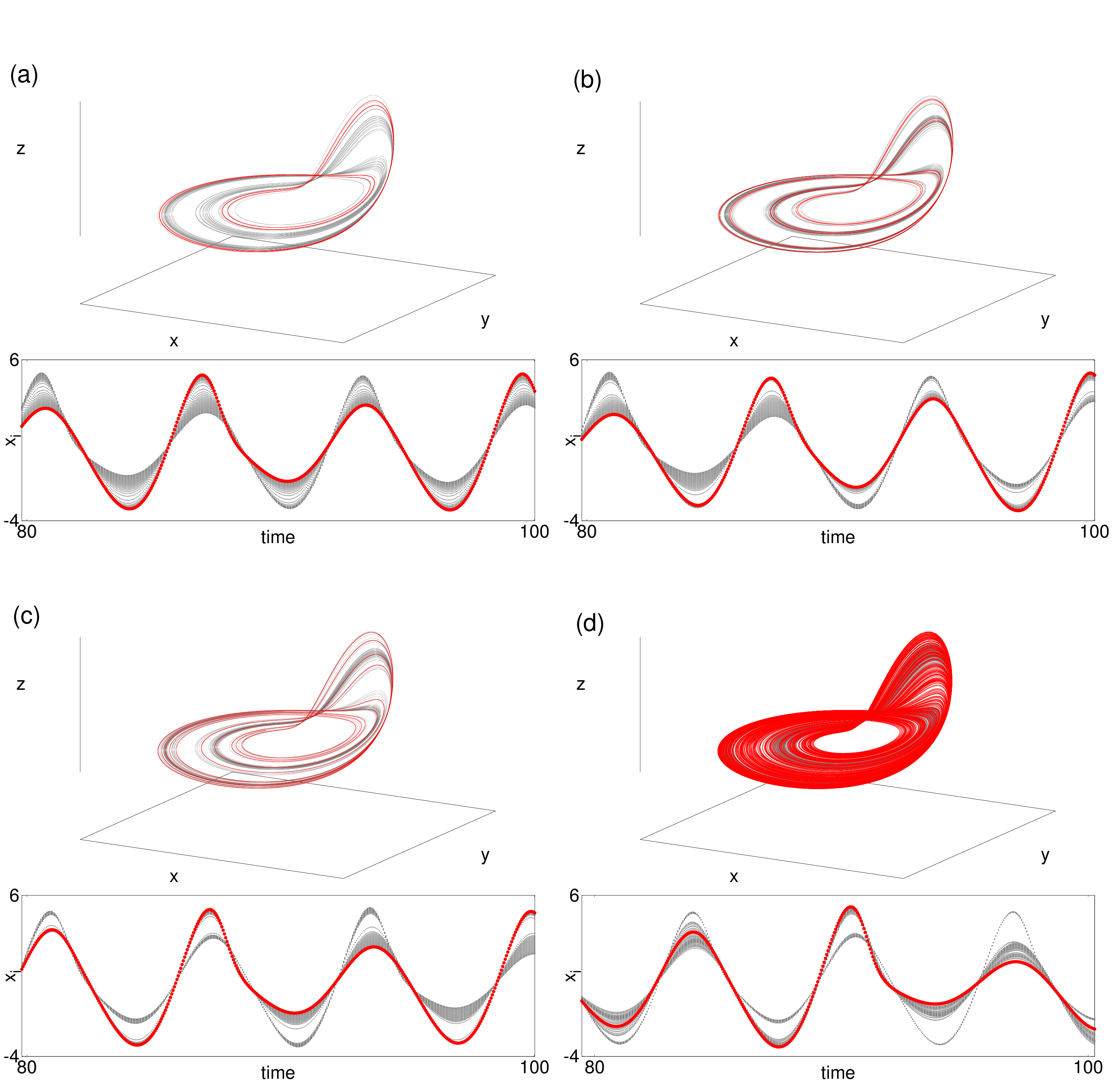}
\caption{(Color online) Phase space trajectories (upper panels) and time series $x_i(t)$, $i=1,2,...,N$, (lower panels) for the
R\"ossler system. The coupling strength is chosen as $\sigma= 0.12,
0.1, 0.09$, and $0.06$ in panels (a) to (d), respectively, corresponding to Fig.~\ref{fig:RoesslerBifScenario}. 
The exemplary time series $x_{50}(t)$ is highlighted in black (red).
The coupling radius is fixed at $r=0.28$. Other parameters as in Fig.~\ref{fig:diagram}.}
\label{fig:Roessler_XoverT}
\end{figure*}

Figures~\ref{fig:RoesslerBifScenario} and \ref{fig:Roessler_XoverT} display the typical scenario of
coherence-incoherence transition for nonlocally coupled R\"ossler systems, similar to the one described above for the
logistic map. In the upper panels of Fig.~\ref{fig:RoesslerBifScenario}, snapshots of the variables $x_i$ are shown.
Here all nodes are shown in gray, and an exemplary node ($i=50$) is
highlighted as a black point. 
Similarly to the coupled maps, the smooth profile of the coherent ($k=1$) solution of
Fig.~\ref{fig:RoesslerBifScenario}(a) breaks up into two clusters in Fig.~\ref{fig:RoesslerBifScenario}(b). With further
decrease of the coupling strength a partially coherent solution appears in the system as shown in
Fig.~\ref{fig:RoesslerBifScenario}(c). Finally we observe a chimera-like state around the upper cluster state in
Fig.~\ref{fig:RoesslerBifScenario}(d), which combines spatially coherent and spatially chaotic parts.  The lower panels
of Fig.~\ref{fig:RoesslerBifScenario} show space-time plots. Here the spatial structure and the breaking up of the
profile is clearly visible. Furthermore one can already infer the temporal behavior that will be discussed
in the following.

Figure~\ref{fig:Roessler_XoverT} shows three-dimensional phase portraits of the $x_i$ variables and 
the superimposed time series of all $x_i(t)$, $i=1,...,N$ of the R\"ossler
systems for the same parameter values as in Fig.~\ref{fig:RoesslerBifScenario}. Again the time series of node $i=50$
is highlighted as black (red). Note that in Figs.~\ref{fig:Roessler_XoverT}(b) and (c) the time series of the points
from the incoherent regions are grouped into two clusters that are formed by the two coherent parts of the solution. In
Fig.~\ref{fig:Roessler_XoverT}(d) one can visualize the chaotic dynamics already in the short time interval shown. In
the time domain one can observe a period-doubling bifurcation in Figs.~\ref{fig:Roessler_XoverT}(a) to
\ref{fig:Roessler_XoverT}(c) eventually leading to chaos in Fig.~\ref{fig:Roessler_XoverT}(d) (lower panels). At
the same time the spatial irregularity becomes more pronounced as described above.

\section{Analytical results}\label{sec:analytical}
After the numerical exploration of the coherence-incoherence transition, we provide some analytical derivations in this Section. We will discuss an approximation for the value of the critical coupling strength at the coherence-incoherence transition and present a scaling of the profiles in different coherence tongues, which holds in the limit $N\rightarrow\infty$.

\subsection{Critical coupling strength}\label{sub:strength}
In the following, we derive an approximation of the critical coupling strength at the transition from coherent to incoherent states, see the line CIB in Fig.~\ref{fig:diagram}(a).  We will restrict the investigation to the logistic map introduced in Sec.~\ref{sub:map}, but a similar argument holds for the R\"ossler system as well.

From a geometrical point of view and in the thermodynamic limit $N\rightarrow \infty$, coherent solutions $z_i^t$ approach a smooth profile  $z^t(x)$ of the spatially continuous version of Eq.~(\ref{eq:map}) given by
\begin{equation}\label{eq:continuous}
z^{t+1}(x)= f\left(z^t(x)\right)+ \frac{\sigma}{2r} \int_{x-r}^{x+r} \left[f\left(z^{t}(y)\right) -f\left(z^{t}(x)\right)\right] dy.
\end{equation}
A transition from coherence to incoherence occurs when the respective solution profile $z^t(x)$ 
becomes discontinuous in some points $x$ of the ring $\mathcal{S}^{1}$.

\begin{figure}[ht!]
\includegraphics[width=\linewidth]{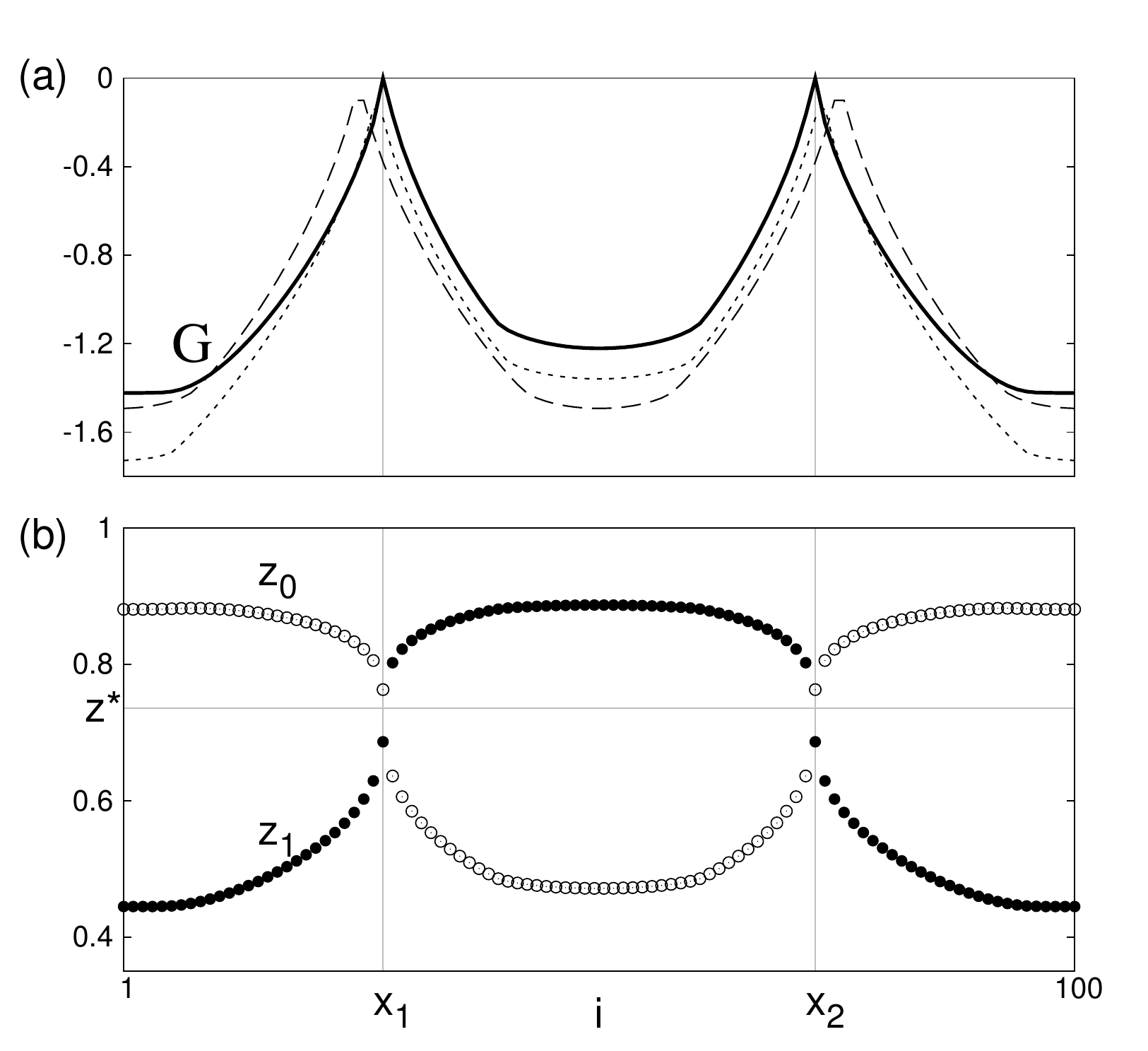}
\caption{(a) Deviation from the incoherence condition~(\ref{eq:cond}) for coupled logistic maps for $\sigma=$ 0.42
(dashed), 0.40919 (solid curve), $\sigma=0.40$ (dotted). (b) Snapshots for even (open dots) and odd (filled dots)
solutions for $\sigma=0.40919$ and $r=0.32$. Other parameters as in Fig.~\ref{fig:diagram}.}
\label{fig:analytic}
\end{figure}

Let us now discuss in more detail this coherence-incoherence bifurcation. Consider a solution of system~(\ref{eq:continuous}) with wave number $k=1$ and period-2 dynamics in time. Hence we can reduce the dynamics by even and odd time steps $z_0(x)$ and $z_1(x)$, respectively. This leads to
\begin{equation}
z_{1-j}(x) = (1-\sigma)f\left(z_j(x)\right) + \frac{\sigma}{2r} \int\limits_{x-r}^{x+r}f\left(z_j(y)\right)dy
\end{equation} 
with $j=0, 1$. Taking the spatial derivative yields
\begin{align}\label{eq:deriv}
z_{1-j}'(x) =& (1-\sigma)f'\left(z_j(x)\right)z_j'(x) \nonumber\\
& + \frac{\sigma}{2r} \left[f\left(z_j(x+r)\right) - f\left(z_j(x-r)\right) \right].
\end{align} 

At the point $x$ where the smooth profile breaks up, the spatial derivative becomes infinite. Considering that $z_0'(x), z_1'(x)$ diverge to infinity, we can neglect the coupling term on the right-hand side of Eq.~(\ref{eq:deriv}). The main contribution comes from the first term.

Multiplying the equations for even and odd time steps we obtain:
\begin{equation}
z_{0}'(x)z_{1}'(x) = \left[ (1-\sigma)^2 f'\left(z_0(x)\right)f'\left(z_1(x)\right) \right] z_{0}'(x)z_{1}'(x),
\end{equation} 
which yields the following condition
\begin{equation}\label{eq:cond}
1 = (1-\sigma)^2 f'\left(z_0(x)\right)f'\left(z_1(x)\right) .
\end{equation} 
Keeping in mind that the local dynamics is governed by the logistic map $f(z)=az(1-z)$ and that its derivative is equal
to $f'(z)=a(1-2z)$, we introduce the function 
\begin{equation}\label{eq:G}
 G(x) = (1-\sigma)^2 a^2 \left(1-2 z_0(x)\right)\left(1- 2 z_1(x)\right) - 1
\end{equation} 
as a deviation from condition~(\ref{eq:cond}).

For fixed system parameters $r$ and $\sigma$ that correspond to the region of coherent solutions with periodic behavior of period $2$ in time, and thus fixing $z_0(x)$ and $z_1(x)$, we can calculate $G(x)$ and determine values $x$, where $G(x) = 0$. The obtained values $x$ are the locations at which the smooth profile break up into two parts. Figure~\ref{fig:analytic} illustrates $G(x)$ for a fixed coupling radius $r=0.32$ for the finite case of 100 elements. Numerical evaluation shows that for a coupling strength $\sigma\approx0.41$ the function $G$ approaches zero in two points. Figure~\ref{fig:analytic}(b) displays the corresponding snapshot of the solution. One can see that the profile breaks up at positions at which $G$ vanishes as indicated by the vertical dashed lines.

To derive an approximation for this coupling strength let  $z^*$  be the fixed point of the local logistic map: $z^*= f(z^*)=a z^*(1-z^*)$, hence $z^*= 1- 1/a$. 
Here, for $a=3.8$ we have $z^*\approx0.737$. Under the assumption $z_0(x)=z_1(x)=z^*$ if $G=0$, we obtain an approximation for $\sigma$:
\begin{align}
G(x)&= \left[ a(1-\sigma)(1-2 z^*) \right]^2 -1 = 0 \\
&\Rightarrow \sigma \approx 1 - \frac{1}{a-2}.
\end{align}
For $a=3.8$ we get $\sigma \approx 0.44$.
 
From the numerical experiments we obtain that the transition from coherence to incoherence in system~(\ref{eq:map}) occurs for a coupling strength $\sigma$ close to $0.4$,  which roughly agrees with our approximation. The deviation is due to
the finite number of nodes in the simulations. The derivation above was presented for 
wavenumber $k=1$. Note
that coherent solutions with higher wave numbers undergo the same bifurcation scenario, with the smooth profile breaking up into $2k$ parts.

\subsection{Scaling of coherent profiles}\label{sub:scaling}
\begin{figure}[ht!]
\includegraphics[width=\linewidth]{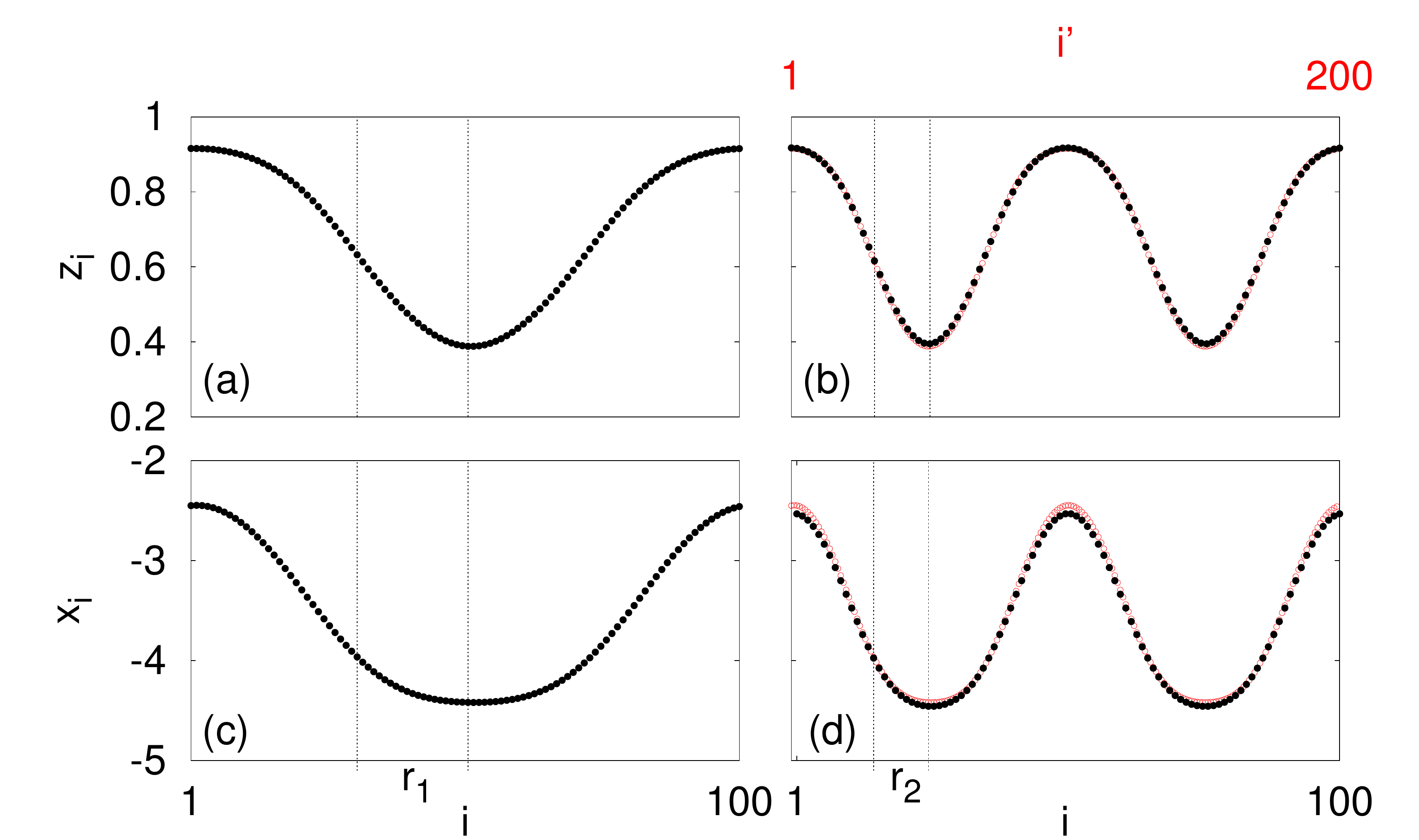}
\caption{(Color online) 
Scaling of coherent profiles: Black points are snapshots of numerical simulations (a),(b) for coupled logistic maps with 
$N=100,$ $\sigma=0.7,$ $r=0.22$ and $r=0.11$, respectively, and (c),(d) for R\"ossler systems with $N=100$,
$\sigma=0.6$, $r=0.12$ and  $r=0.06$, respectively. Gray (red) points in panels (b),(d) show rescaled $k=1$ solutions
from panels (a) and (c). Other parameters as in Fig.~\ref{fig:diagram}.}
\label{fig:kscale}
\end{figure}

As a second analytic result we will derive in the following a scaling relation that allows a mapping of coherent profiles in different regimes of coherence. These regimes were depicted in Fig.~\ref{fig:diagram} by shaded (color coded) tongues labeled $k=1,2,\dots$. Our considerations apply to rings of oscillators in the thermodynamic limit $N\to \infty$, $N$ being the number of coupled oscillators. A comparison with numerical simulation shows that for sufficiently large finite $N$ the result holds in good approximation. For notational convenience, we consider the case of time-continuous systems, but our findings are also valid for maps as we will show later (Fig.~\ref{fig:kscale}).

The network model of a ring of $N$ linearly coupled oscillators, each coupled to its $P$ nearest neighbors,
\begin{equation}
\dot{{\bf x}}_i = {\bf f}\left( {\bf x}_i\right) + \frac{\sigma}{2P} \sum_{j=i-P}^{i+P} \left( {\bf H} {\bf x}_j - {\bf H} {\bf x}_i \right)
\label{model}
\end{equation}
becomes, in the limit $N \to \infty$,
\begin{equation}
\dot{{\bf x}}(u) = {\bf f}({\bf x}(u)) + \frac{\sigma}{2r} \int_{u-r}^{u+r} \left[{\bf H} {\bf x}(v) - {\bf H} {\bf x}(u) \right] \mbox{d}v
\label{modeltherm}
\end{equation}
with coupling range $r$ and coupling strength $\sigma$ as used in the previous sections. $u\in\left[ 0,1 \right]$ denotes the spatial position on the ring. Compare Eq.~(\ref{eq:continuous}) for the spatially continuous version of time-discrete maps. The coupling scheme is determined by the matrix ${\bf H}$. In the case of the R\"ossler system introduced by Eqs.~(\ref{eq:roessler}), ${\bf H}$ is the identity matrix and the vector describing each node is given by ${\bf x}_i=(x_i,y_i,z_i)$. Considering the logistic maps of Eq.~(\ref{eq:map}), the coupling term  ${\bf H}{\bf x}_j$ changes to $f(z_j^t)$, but the following argument still holds (Fig.~\ref{fig:kscale}(a),(b)).  

In short, we will show that for fixed $\sigma$ the spatially $k_1$-periodic dynamics of an oscillator at the position $u$ with coupling range $r_1$ is identical to that at a rescaled position $u k_1/k_2$ of the corresponding $k_2$-periodic state in a network with coupling range $r_2= r_1 k_1/k_2$.
From this we can conjecture to find a $k_2$-periodic state at $r_2$, if a stable $k_1$-periodic state is found at $r_1$, such that $k_ir_i=const.$

Assume that a state ${\bf x}_s(u)$ with spatial period $k_1$ exists. For simplicity and notational convenience consider $k_1=1$. This state evolves according to
\begin{equation}
\dot{{\bf x}}_s(u) = {\bf f}\left({\bf x}_s(u)\right) + \frac{\sigma}{2r} \int_{u-r}^{u+r} \left[ {\bf H} 
    {\bf x}_s(v) - {\bf H} {\bf x}_s(u) \right] \mbox{d}v.
\label{unscaled}
\end{equation}
In order to compare another $k$-periodic state ${\bf x}_k$ with the state ${\bf x}_s$,
in the way described above, we assume

\begin{equation}\label{periodrel}
 {\bf x}_s(u) = {\bf x}_k\left(\frac{u+m}{k}\right)
\end{equation}
with $m=0,1,2,\dots,k-1$. Vice versa, scaling the spatial coordinate of ${\bf x}_k(u)$ with $u\in[0,1]$, as well as $r$, by a factor $k^{-1}$ leads to
\begin{align}
\dot{{\bf x}}_k(k^{-1}u) = &{\bf f}\left({\bf x}_k(k^{-1}u)\right) \nonumber\\ 
   &+ \frac{k\sigma}{2r} \int_{k^{-1}(u-r)}^{k^{-1}(u+r)} 
	\left[ {\bf H} {\bf x}_k(v) - {\bf H} {\bf x}_k(k^{-1}u)  \right] \mbox{d}v.
\end{align}
Substituting $v=k^{-1}w$ gives
\begin{align}
\dot{{\bf x}}_k(k^{-1}u) = &{\bf f}\left({\bf x}_k(k^{-1}u)\right) \nonumber\\ 
   &+ \frac{\sigma}{2r} \int_{u-r}^{u+r} 
	\left[ {\bf H} {\bf x}_k(k^{-1}w) - {\bf H} {\bf x}_k(k^{-1}u)  \right]\; \mbox{d}w
\label{scaled}
\end{align}
and, using Eq.~\eqref{periodrel}, results in

\begin{equation}
\dot{{\bf x}}_k(k^{-1}u) 
= {\bf f}\left({\bf x}_s(u)\right) + \frac{\sigma}{2r} \int_{u-r}^{u+r} 
	\left[ {\bf H} {\bf x}_s(w) - {\bf H} {\bf x}_s(u)  \right] \mbox{d}w.
\label{scaleeq}
\end{equation}
Comparing the right-hand side of Eq~(\ref{scaleeq}) to Eq.~(\ref{unscaled}), we find $\dot{{\bf x}}_k(k^{-1}u)=\dot{{\bf x}}_s(u)$, if the coupling radius $r$ is appropriately rescaled by $k^{-1}$. 

This derivation shows that one can recover the coupling parameters leading to a state with wave number $k_2$ from the
parameters of a $k_1$ state. This finding is illustrated in Fig.~\ref{fig:kscale} for $k_1=1$ and $k_2=2$ by
corresponding snapshots of 100 logistic maps and 100 R\"ossler system in panels (a),(b) and (c),(d), respectively.
Already for $N=100$ there is strong agreement between the numerically obtained and scaled profiles. In the case of
logistic maps (R\"ossler systems), we choose the coupling strength as $\sigma=0.7$ ($0.6$) and the coupling radius as
$r=0.22$ ($0.12$) in Fig.~\ref{fig:kscale}(a) (Fig.~\ref{fig:kscale}(c)) and, according to the scaling relation, as
$r=0.11$ ($0.06$) for the $k_2=2$ profile in Fig.~\ref{fig:kscale}(b) (Fig.~\ref{fig:kscale}(d)). The gray (red) dots
in Figs.~\ref{fig:kscale}(b) and (d) depict the scaled profile of Fig.~\ref{fig:kscale}(a) and (b), respectively. 

\begin{figure}[ht!]
\includegraphics[width=\linewidth]{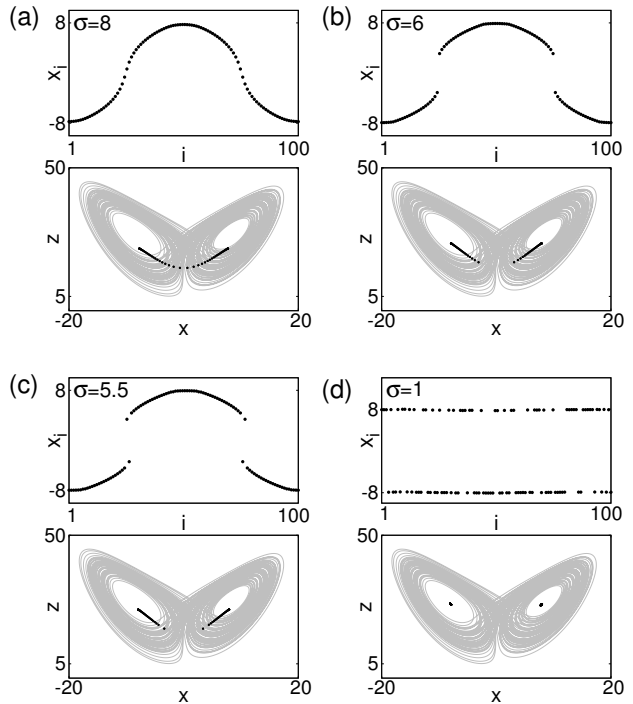}
\caption{Snapshots of nonlocally coupled Lorenz systems~(\ref{eq:lorenz}) for different values of coupling strength $\sigma=8, 6, 5.5$, and $1$ in panels (a) to (d), respectively. The coupling radius is fixed at $r=0.3$ and the number of nodes is $N=100$. Parameters of the Lorenz system: $\varsigma=10,$ $\rho=28,$ $\beta=8/3$.}
\label{Fig:LorenzSnapShots}
\end{figure}

\section{Lorenz model}\label{sec:lorenz}
To investigate how universal the results obtained so far are, we provide one more example of a chaotic time-continuous system that exhibits the coherence-incoherence bifurcation scenario described above. Figure~\ref{Fig:LorenzSnapShots} displays the coherence-incoherence transition for nonlocally coupled Lorenz systems:
\begin{equation}\label{eq:lorenz}
\begin{array}{l}
\dot{x}_i = \varsigma(y_i-x_i) + \dfrac{\sigma}{2P} \sum\limits_{j=i-P}^{i+P} \left( x_j - x_i \right), \\
\dot{y}_i = x_i (\rho -z_i) - y_i  + \dfrac{\sigma}{2P} \sum\limits_{j=i-P}^{i+P} \left( y_j - y_i \right), \\
\dot{z}_i = x_i y_i - \beta z_i +\dfrac{\sigma}{2P} \sum\limits_{j=i-P}^{i+P} \left( z_j - z_i \right), 
\end{array}
\end{equation} 
where the local parameters are fixed at the values $\varsigma=10,$ $\rho=28,$ $\beta=8/3$.
Fig.~\ref{Fig:LorenzSnapShots} reveals a striking similarity with the sequences of snapshots for nonlocally coupled
chaotic logistic maps (Fig.~\ref{fig:maps_snap}) and R\"ossler systems (Fig.~\ref{fig:RoesslerBifScenario}). For fixed
coupling radius and with decreasing coupling strength the smooth profile breaks up into two parts
(Fig.~\ref{Fig:LorenzSnapShots}(b)). Further decrease of the coupling strength leads to the appearance and growth of
incoherent parts, repeating again the typical scenario of the coherence-incoherence bifurcation described above.

Additional numerical evidence shows that system~(\ref{eq:lorenz}) is characterized by high multistability involving
different types of solutions. There are no clearly separated tongues in the $(r,\sigma)$-parameter plane
present as in logistic maps and R\"ossler systems, but there is strong overlap of the regimes of different wave
numbers $k=1,2,3,\dots $ and the region of spatially coherent synchronized chaotic dynamics ($k=0$). This is due to the
completely different local dynamics, which is characterized by a double  scroll and two symmetric saddle-foci. The
coupling induces stable spatially coherent steady states in the form of wavelike profiles with wave numbers $k=1,2,3,\dots$
(Fig.~\ref{Fig:LorenzSnapShots}(a)).

\section{Conclusion}\label{sec:conclusion}
By both numerical and analytical means we have investigated the transition from coherence to incoherence in coupled chaotic systems. In dependence upon the coupling strength and coupling range we have observed a combination of period-adding in space and period-doubling in time for nonlocally coupled chaotic maps and R\"ossler systems. 
An inspection of the spatial profiles reveals a coherence-incoherence bifurcation, where the smooth coherent profile of
the solution breaks up into parts and gives rise to partially coherent solutions and chimera states. Upon further
decrease of the coupling strength completely incoherent states are found as a sign of fully developed spatial chaos. 
A similar transition from coherence to incoherence is found for nonlocally coupled Lorenz systems. Our findings hold for both time-discrete and time-continuous systems indicating the universality of the transition. 

In conclusion, we have analyzed in detail the transition between spatial coherence and incoherence that appears in
different systems with nonlocal couplings, and we find similar bifurcation scenario for nonlocally coupled logistic
maps, R\"ossler and Lorenz systems. Beyond numerical investigations, we have determined analytically the critical
coupling strength for this bifurcation, and established a scaling relation for profiles with different wave numbers.

\begin{acknowledgments}
 IO and PH acknowledge support by the BMBF under the grant no. 01GQ1001B (\textit{F\"orderkennzeichen}). 
 ES and YM acknowledge support by Deutsche Forschungsgemeinschaft in the framework of SFB910. We thank Edward Ott for
stimulating discussions.
\end{acknowledgments}


\end{document}